\documentclass[aps,prb,twocolumn,superscriptaddress,longbibliography]{revtex4-2}

\usepackage{amsmath,amssymb,bm}
\usepackage{graphicx}
\usepackage{hyperref}
\usepackage{xcolor}
\usepackage{longtable}
\usepackage{booktabs}
\begin{document}

\title{Breakdown of the periodic potential ansatz in correlated electron systems}

\author{Wouter Montfrooij}
\affiliation{Department of Physics and Astronomy, University of Missouri, Columbia, Missouri 65211, USA }

\begin{abstract}
Our electronic structure theory for crystalline solids is commonly built on the periodic potential assumption
$V(\mathbf r)=V(\mathbf r+\mathbf R)$ for every lattice translation $\mathbf R$, enabling Bloch eigenstates,
crystal momentum as a good quantum number, and the standard quasiparticle-based description of the behavior of metals.
Because the zero-point motion of the ions, however, in correlated electron systems the electronic environment experienced by an itinerant electron is neither
static nor self-averaging at the single-particle level, even in perfectly stoichiometric crystals, leading to a distribution of local Kondo scales that spans two orders of magnitude in temperature. We discuss, through a comparison between uniform scenarios and one that breaks with perfect lattice translational symmetry, how incorporating this distribution yields a unified description for all heavy-fermion systems at the quantum critical point.
\end{abstract}

\maketitle

\section{Introduction}

The theoretical description of electrons in solids rests on the intuitively obvious assumption that the effective potential experienced by an electron is periodic with the periodicity $\mathbf R$ of the lattice:
\begin{equation}
V(\mathbf r) = V(\mathbf r + \mathbf R).
\end{equation}
This assumption, based on the almost negligible amplitude of the ionic zero-point motion, underlies Bloch’s theorem \cite{Bloch1929}, the existence of electronic band structures, the concept of a Fermi surface, Landau’s Fermi-liquid theory \cite{Landau1957}, and essentially all modern electronic structure methods \cite{KohnSham1965}. Its empirical success across metals, semiconductors, and insulators has made it one of the most foundational organizing principles of condensed-matter physics.

At the same time, quantum critical heavy-fermion systems display phenomena that have resisted a unified explanation for decades \cite{Stewart2001,Miranda2005,Stewart2006,Lohneysen2007}. Non-Fermi-liquid thermodynamics, anomalous magnetic response, linear-temperature resistivity, multiple low-energy scales, and sharp features in transport coexist in ways that cannot readily be captured by conventional interpretations. Numerous theoretical frameworks have been proposed, including Hertz--Millis--Moriya (HMM) quantum criticality \cite{Hertz1976,Millis1993,Moriya1985}, Kondo breakdown scenarios, local quantum criticality (LQT), and Fermi-surface reconstructions \cite{Si2001,Coleman2001}. Yet despite their sophistication, these approaches have (thus far) failed to produce a cohesive description that accounts simultaneously for all experimental findings.

Here we discuss that the central difficulty lies not in the complexity of the underlying interactions, but in the uniformity imposed at the outset. We argue that in quantum critical heavy-fermion systems the periodic potential ansatz fails in a fundamental way, even in perfectly stoichiometric crystals. The failure is not driven by chemical inhomogeneity or sample imperfections, but arises inevitably from basic quantum-mechanical effects: the zero-point motion of the ions and the extreme sensitivity of the Kondo coupling to interatomic spacing. Together, these effects generate an intrinsic, broad distribution of local Kondo temperatures that destroys spatial uniformity so that the effective static one-body potential seen by electrons on electronic timescales is no longer periodic.

We review how this intrinsic nonuniformity provides a natural microscopic origin for the emergence of magnetic clusters and superspin physics, leading to a percolative description of quantum critical behavior. This framework quantitatively accounts for the thermodynamic, magnetic, and transport properties of several canonical heavy-fermion systems, as we illustrate by revisiting the heavily-studied \cite{Custers2003,Paschen2004,Gegenwart2005,Custers2010,Si2024} quantum critical compound YbRh$_2$Si$_2$. 

The paper is organized as follows. In Sec.~II we present the microscopic origin of the breakdown of the periodic potential, followed by the emergence of clusters and superspin physics. In Sec.~III we summarize the low-temperature phenomenology that any viable theory must explain and we apply the cluster scenario framework to YbRh$_2$Si$_2$ and related compounds. In Sec.~IV we list how experimental observations are interpreted in the cluster scenario. In Sec.~V we review organizing principles against the backdrop of the chaotic groundstate. In Sec.~VI we scrutinize the Doniach phase diagram and we conclude in Sec.~VII.

\section{Microscopic origin of the Kondo distribution and its consequences}
In view of the prevalence of assuming a uniform heavy-fermion ground state from the outset (Ref. \cite{Si2024} and almost all references therein), we review the theoretical underpinning behind the failure of the (effective static one-body) periodic potential ansatz and revisit experimental evidence for quantum critical heavy-fermion systems that demonstrates that the low-temperature behavior of these systems are much easier interpreted in terms of the consequences of a distribution of Kondo energy scales. 

The starting assumption of essentially all microscopic theories of heavy-fermion metals is that the electronic environment is spatially uniform, so that a single Kondo energy scale \cite{Kondo1964} characterizes the entire sample. For electronic structure calculations this translates to assuming a perfectly periodic potential $V(\mathbf r)=V(\mathbf r+\mathbf R)$ \cite{AshcroftMermin}, and in the language of many-body theory it implies a uniform heavy-fermion ground state. However, this assumption fails for quantum critical systems.

\subsection{Zero-point motion and fluctuations of local couplings}

The Kondo scale \cite{Kondo1964} depends exponentially on the hybridization between localized $f$ electrons and conduction electrons, which itself depends sensitively on the instantaneous interionic separation \cite{Endstra1993,Sachdev1999}. For a Kondo lattice one may write \cite{Anderson1961,Hewson1993}
\begin{equation}
k_B T_K = D \exp\!\left[-\frac{1}{J\rho(E_F)}\right]
       = D \exp\!\left[-A r^{12}\right],
\end{equation}
where $D$ is the conduction bandwidth, $r$ the relevant interionic distance, and the second equality (where we have combined all constants into one constant $A$) follows from the known $r$-dependence of the $f$--$d$ hybridization \cite{Endstra1993}. Differentiating and eliminating the unknown constant $A$ in favor of the Kondo scale yields \cite{Bretana2021}
\begin{equation}
\frac{dT_K}{T_K}
   = 12 \ln\!\left(\frac{D}{k_B T_K}\right)\frac{dr}{r}.
\label{TKspread}
\end{equation}

For heavy-fermion systems, typical values are $D\sim$ 0.3--1 eV and $T_K\sim20$--$80$~K, implying $\ln(D/k_BT_K)\approx4$--$6$. Diffraction measurements show that at $T=0$ the zero-point motion of ions produces small relative displacements $dr/r \approx 1$--$2\%$ \cite{Dudschus2019,Bretana2021}. However, inserting these numbers into Eq.~(\ref{TKspread}), we find
\begin{equation}
\frac{dT_K}{T_K} \approx 50\% - 100\%.
\end{equation}
Thus, quantum zero-point motion alone produces an order-unity distribution of local Kondo temperatures, even in a perfectly stoichiometric crystal. Note that these spatial fluctuations of the local Kondo scale cannot be treated as a simple statistical averaging that would restore an effectively periodic potential. The characteristic time scale for electronic motion is many orders of magnitude shorter than that associated with ionic zero-point motion \cite{Born1927}. On the time scale relevant for electronic coherence, the ionic configuration is effectively frozen, and the electrons experience a quasi-static spatial landscape of local couplings rather than a time-averaged uniform background. Over experimental time scales, transport and thermodynamic measurements may sample many such configurations and therefore exhibit apparent averaging. However, the formation of Bloch states and long-lived quasiparticles requires spatial coherence on the electronic time scale, not on the measurement time scale. It is this separation of time scales that prevents the system from self-averaging into an effectively periodic one-body potential, despite the absence of chemical disorder and despite the dynamical nature of the underlying lattice fluctuations. In other words, while the lattice motion is slow enough to justify an adiabatic separation of electronic and ionic degrees of freedom, it is not fast enough to restore spatial self-averaging on the coherence scale required to construct low-energy Bloch-band quasiparticles.

\begin{figure}
\begin{center} 
\includegraphics*[viewport=110 145 580 410,width=85mm,clip]{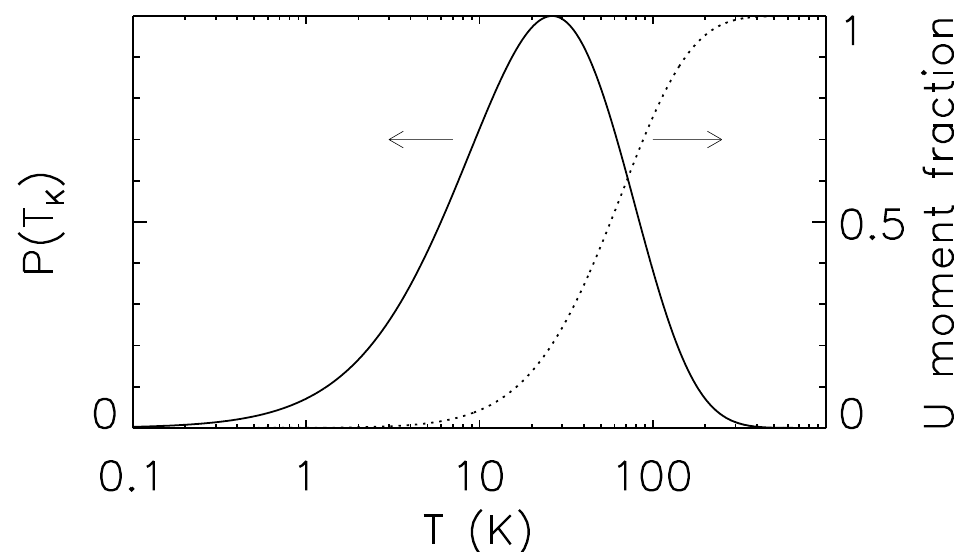}
\end{center}
	\caption{Figure adapted from Ref. \cite{Bretana2023}. The instantaneous distribution of Kondo energy scales for stoichiometric URu$_2$Si$_2$ (solid line, left axis) and corresponding fraction of unshielded moments (dotted curve, right axis) calculated from eqn \ref{TKspread}. The three parameters used were an average $T_K=70$ K estimated from the resistivity coherence temperature, a bandwidth $D=0.4$~eV from band-structure estimates \cite{Elgazzar2009}, and a zero-point displacement of $0.006$~nm obtained from EXAFS experiments \cite{Dudschus2019}.}
	\label{kondoplot}
\end{figure}

This amplification is unavoidable and has profound consequences. Fig. \ref{kondoplot} shows the resulting instantaneous distribution of Kondo shielding temperatures for stoichiometric URu$_2$Si$_2$, obtained by combining a Gaussian distribution of interatomic separations (set by the Debye--Waller factor) with probability conservation $P(T_K)\,dT_K = P(r)\,dr$ \cite{Bretana2023}. Using an average $T_K=70$ K estimated from the resistivity coherence temperature, a bandwidth $D=0.4$~eV from band-structure estimates \cite{Elgazzar2009}, and a zero-point displacement of $0.006$~nm from EXAFS measurements for the U-Ru distance \cite{Dudschus2019}, we find a distribution of Kondo temperatures spanning nearly two orders of magnitude. Thus, a mere two-percent uncertainty in equilibrium atomic positions generates a macroscopically nonuniform electronic ground state in this stoichiometric system, with the distribution width only enhanced for systems with a lower average Kondo scale.

Because electronic time scales are much faster than ionic ones \cite{Born1927}, the electrons experience this distribution as effectively static: the time-tested adiabatic approximation of phonon physics applies. The only way this intrinsic nonuniformity could be ignored would be if the adiabatic approximation were to fail at low temperatures, for which there exists no experimental evidence. The consequence is far-reaching: even the cleanest heavy-fermion metal cannot be described by a spatially uniform heavy-fermion fluid, yet this is exactly how we describe heavy-fermion systems in the literature. However, the periodic one-body potential ceases to exist in any meaningful sense at low energies: the fundamental assumption that underlies Bloch's theorem \cite{Bloch1929} and much of modern electronic structure theory is violated by the quantum mechanics of the lattice itself.

\subsection{Spontaneous emergence of magnetic clusters}

The intrinsic distribution of Kondo temperatures described above has an immediate and unavoidable consequence: upon cooling, magnetic ions cross their local Kondo-shielding threshold at different temperatures and it becomes energetically favored for those moments to be shielded. Magnetic moments therefore disappear in a spatially inhomogeneous fashion, producing isolated islands of unshielded moments embedded in a background of Kondo-singlet regions. This process generates a collection of finite magnetic clusters as well as a lattice-spanning infinite cluster \cite{StaufferAharony,Fayfar2021a}.

Once a finite cluster becomes isolated, quantum mechanical finite-size effects dominate its internal dynamics \cite{Heitmann2014}. The energy cost disturbing the alignment of moments inside such a cluster grows as the cluster shrinks since the wavelength of disordering fluctuations must match the size of the cluster: the smaller the cluster, the shorter the minimal wavelength and the higher the corresponding energy required to impose such a disordering fluctuation. As a result, the Ruderman-Kittel-Kasuya-Yosida (RKKY) interaction \cite{Ruderman1954,Kasuya1956,Yosida1957} forces the moments inside each finite cluster to align with their neighbors, as observed experimentally  \cite{Montfrooij2007,Montfrooij2019,Bretana2021}. When this happens, Kondo shielding becomes strongly suppressed, since the Kondo mechanism relies on spin-flip processes that are energetically unfavorable in an ordered local environment \cite{Kondo1964}. Each cluster therefore behaves as a single large magnetic object — a \emph{superspin} — whose magnitude is set by the number and arrangement of the uncompensated moments within the cluster. Note that independent of whether the moments on these clusters align with their neighbors or not, the spontaneous fragmentation of the lattice into a collection of clusters upon cooling will take place and the relation $V(\mathbf r)=V(\mathbf r+\mathbf R)$ for the effective static potential breaks down. We provide experimental evidence for superspins materializing in the next subsection.

The percolation scenario (or cluster scenario) described above has been analyzed by means of Monte Carlo computer simulations for various 2D and 3D lattice structures, both for isolated clusters being protected from Kondo shielding or remaining susceptible to it. We refer to the literature for the details  \cite{Fayfar2021a,Fayfar2021b}. These simulations yield the full distribution of cluster sizes and superspin moments near the percolation threshold, and provide the quantitative input required to compute the temperature dependence of the magnetic susceptibility and specific heat of a system governed by this physics. The percolation scenario was tested against the available experimental data in a host of systems and yielded quantitative agreement both in chemically substituted as well as in stoichiometric quantum critical compounds \cite{Bretana2021,Bretana2023}. In Section III we summarize these findings emphasizing the canonical system YbRh$_2$Si$_2$ as an illustrative example so that we can do a side by side comparison of interpretations based on our cluster scenario and those based on a scenario assuming uniformity in Section IV.

\subsection{Evidence for cluster superspins}
In order for the quantitative predictions of the cluster scenario to stand up to scrutiny, it is necessary for the moments on isolated clusters to indeed align with their neighbors. From a theoretical point of view, it is clear that a truly isolated cluster at low temperature must have its moments aligned because of finite size effects. However, these clusters are embedded in a lattice with long-range RKKY interaction \cite{Ruderman1954,Kasuya1956,Yosida1957}  so the issue to be considered is how effectively isolated are these clusters from each other as well as from the infinite cluster. 

A cluster with all its moments lined up can readily be identified by means of neutron scattering in a non-cubic system as the appearance of a superspin \cite{Montfrooij2007} comes with a prediction unique to the percolation scenario. In percolation theory, the clusters form according to random spatial patterns, not related to the lattice symmetry such as whether the lattice is body-centered cubic or body-centered tetragonal. Therefore, provided the moments in a cluster line up, the number of moments correlated with each other should be independent of crystallographic direction. Thus, even though we would only see short-range order representing the spatial extent of the cluster, the correlation length of this short-range order is not determined by the interaction strength of the moments along different directions and we can expect to observe cubic symmetry in a tetragonal system. This in contrast to incipient long-range order where correlation lengths reflect the symmetry of the lattice. In addition, upon cooling more clusters should form, and additional short-range order materializes without affecting short-range order already present at the higher temperatures. 

The neutron scattering data on chemically doped Ce(Ru$_{0.24}$Fe$_{0.76}$)$_2$Ge$_2$  \cite{Montfrooij2007} and on stoichiometric CeRu$_2$Si$_2$ \cite{Kadowaki2006,Bretana2021} are fully in line with the central predictions of the cluster scenario: the number of correlated moments is dependent of crystallographic direction and on cooling, and scattering intensity piles up on top of existing intensity. However, one might argue that this independence has only been investigated for two systems, and these two systems are very close in crystal structure. The reason for this paucity of data is that experiments on quantum critical systems simply  have not been set up with this type of measurements in mind, not because the measurements are intrinsically difficult. However, there exists another piece of independent evidence for the presence of superspins based in thermodynamics.

\begin{figure}
\begin{center} 
\includegraphics*[viewport=30 105 280 300,width=85mm,clip]{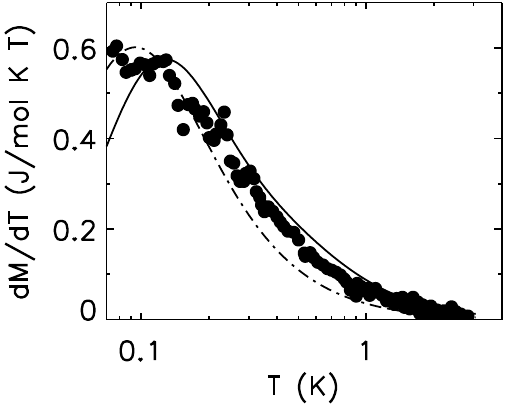}
\end{center}
	\caption{dM/dT data (filled symbols) for YbRh$_2$Si$_2$ digitized from Tokiwa \textit{et al.}\cite{Tokiwa2010}  measured for B= 0.06T applied in the easy plane, corresponding to the critical field $B_c$. The solid curve is the prediction for a collection of superspins at the percolation threshold for a 2-level system with $E= \pm \mu_{\text{eff}}\mu_B B$ for $ \mu_{\text{eff}}$= 1.74 with clusters of size 2 having their AF simulated superspin replaced by that corresponding to an ordering wave vector $\bold{Q}$= (0.14,0.14,0) \cite{Stock2012}, the dotted curve is for clusters upto size 5 having their AF moments corrected. Note that no vertical or horizontal adjustment factors were applied in the comparison between simulation and experiment. Had the simulated clusters not acquired a superspin, then the simulated curves would have been raised by a factor $\sim$20.}
	\label{tokiwaplot}
\end{figure}

For a 2-level system with magnetic moment $\mu$, we have an exact expression for the magnetization $M$ as a function of applied field. In 2010, Tokiwa \textit{et al.} \cite{Tokiwa2010} measured d$M$/d$T$  for YbRh$_2$Si$_2$ in a magnetic field corresponding to the order-disorder transition at $T$= 0. We reproduce their data in Fig. \ref{tokiwaplot}. For a fluctuating entity in a 2-level system, d$M$/d$T$ peaks at 0.8336 $\mu \mu_B B= k_BT$ with a peak height of $k_B/B$. Thus, the peak height is independent of the moment value, as expected given that 

\[
\left(\frac{\partial M}{\partial T}\right)_B
=
\left(\frac{\partial S}{\partial B}\right)_T.
\]

Therefore, a collection of fluctuating entities has a peak height corresponding to the number of fluctuating entities. This allows for a direct distinction between a cluster where all the moments are locked in to form a superspin (a single fluctuating entity) versus a cluster where all the moments are free to fluctuate out of lockstep. At the percolation threshold, the difference between the two options is a factor of about 20; this factor reflects the average number of moments contained in a cluster (of size 2 or larger) at the percolation threshold.  We show the cluster prediction in Fig. \ref{tokiwaplot}, and as is clear from the figure, the numerical agreement is good.

The agreement in $dM/dT$ is particularly significant because the comparison is only weakly sensitive to the microscopic moment size  $ \mu_{\text{eff}}$ of an individual Yb ion as this would only correspond to a scale factor in the temperature axis, but not in the  $dM/dT$ axis. We discuss how we determined $ \mu_{\text{eff}}$  in the next section. Clearly, the measured data and the cluster prediction based on the superspin assumption do not differ by a factor of $\sim$20. This provides strong evidence that the moments within an isolated cluster do not fluctuate independently, but instead align and fluctuate collectively as a single superspin. The factor of 20 is also relevant when we discuss Hall effect measurements in Section IV. Given the neutron scattering results and the magnetization data, we assume in the following that the unavoidable fragmentation of the magnetic lattice into clusters is accompanied by a superspin formation on isolated clusters.

\section{Experimental results for quantum critical systems}

\subsection{What must be explained in quantum critical systems}

Quantum critical heavy-fermion systems display a consistent and well-documented set of low-temperature phenomena \cite{Stewart2001,Miranda2005,Stewart2006} that any theoretical description must address simultaneously. These include a strongly enhanced or divergent electronic specific heat coefficient $c/T$, anomalous magnetic susceptibility, non-Fermi-liquid electrical resistivity often approaching a linear temperature dependence, the emergence of multiple low-energy scales in applied magnetic field \cite{Custers2010}, sharp features in transport coefficients such as the Hall effect \cite{Paschen2004} and the emergence of hyperscaling where temperature is the only energy scale \cite{Aronson1995,Schroder2000,Montfrooij2003}, implying Planckian dynamics.

In addition, different experimental probes frequently infer markedly different effective magnetic moments for the same material \cite{Takahashi2003}, and the characteristic temperature scales extracted from thermodynamic, magnetic, and transport measurements often do not coincide. These observations are not isolated anomalies but constitute the generic phenomenology of quantum critical heavy-fermion compounds. The challenge is therefore not to explain individual measurements in isolation, but to provide a unified description capable of accommodating all of these features within a single physical framework.

Most existing theoretical approaches to quantum criticality in heavy-fermion systems begin by assuming a spatially uniform lattice. Within this framework, the low-energy excitations are described as quasiparticles defined over the entire crystal, and deviations from Fermi-liquid behavior are attributed to collective fluctuations of this homogeneous state. 

However, the experimental phenomenology regarding the coexistence of broad thermodynamic crossovers with sharp transport features \cite{Paschen2004}, the presence of multiple distinct energy scales \cite{Custers2010}, the emergence of $E/T$-scaling \cite{Aronson1995,Schroder2000}, as well as the conflicting estimates of magnetic moments from different probes all require additional mechanisms when interpreted within a spatially uniform picture. Notwithstanding these difficulties, two qualitative, self-consistent scenarios have been developed that we refer to as the Hertz-Millis-Moriya scenario (HMM) \cite{Hertz1976,Millis1993,Moriya1985} and the Local Quantum Criticality (LQT) scenario\cite{Si2001,Paschen2004,Si2024}. We assume the reader is familiar (enough) with these two scenarios and that our brief summary in Section IV.a will be sufficient.

While these two scenarios are very distinct from each other, they have in common that they are both based around a uniform ground state of the critical system. In the following subsection we show that abandoning the principle of uniformity in favor of the cluster scenario simplifies the interpretation of the experimental results on quantum critical systems, providing us with a third scenario for describing quantum critical physics. The main arguments in that section have already been published \cite{Bretana2021,Bretana2023}, in here they are repeated and expanded upon for the sake of completeness and to facilitate the direct comparison between the various scenarios in section IV.

\subsection{Cluster scenario applied to quantum critical systems}

The emergence of magnetic clusters in quantum critical systems was observed by means of neutron scattering experiments on chemically doped Ce(Ru$_{0.24}$Fe$_{0.76}$)$_2$Ge$_2$ where clusters were seen to form \cite{Montfrooij2007} below the average Kondo temperature, with an increasing number appearing upon further lowering of the temperature \cite{Montfrooij2019}. Experiments revealed that  the magnetic moments of the ions making up these clusters were aligned with their neighbors, and that these moments were protected from Kondo shielding \cite{Heitmann2016,Montfrooij2019}, as discussed in Section II.C. 

Later work demonstrated that an identical fragmentation into magnetic clusters also took place \cite{Bretana2021} in stoichiometric quantum critical CeRu$_2$Si$_2$: both the chemically disordered as well as the stoichiometric system displayed an identical distribution of magnetic clusters. In addition, re-analyzing magnetic susceptibility measurements on CeRu$_2$Si$_2$ using the presence of magnetic clusters revealed that the disparate values for the average magnetic moment (three orders of magnitude) that had been inferred \cite{Takahashi2003} from various measurements to (largely) be a manifestation of the underlying distribution of Kondo energy scales.

Recently it was shown that the low-temperature behavior of the canonical stoichiometric heavy-fermion systems UCu$_4$Pd \cite{Bernal1995,Vollmer2000},  CeCu$_6$ \cite{Schroder2000,Tsujii2000}, and  YbRh$_2$Si$_2$ \cite{Custers2003,Gegenwart2005,Custers2010} close to their quantum critical points could also be fully understood \cite{Bretana2023} (in a quantitative manner) as the response of a collection of magnetic clusters without invoking the more involved interpretations put forward in the literature. 

\begin{figure}[t]
\begin{center} 
\includegraphics*[viewport=40 108 545 510,width=85mm,clip]{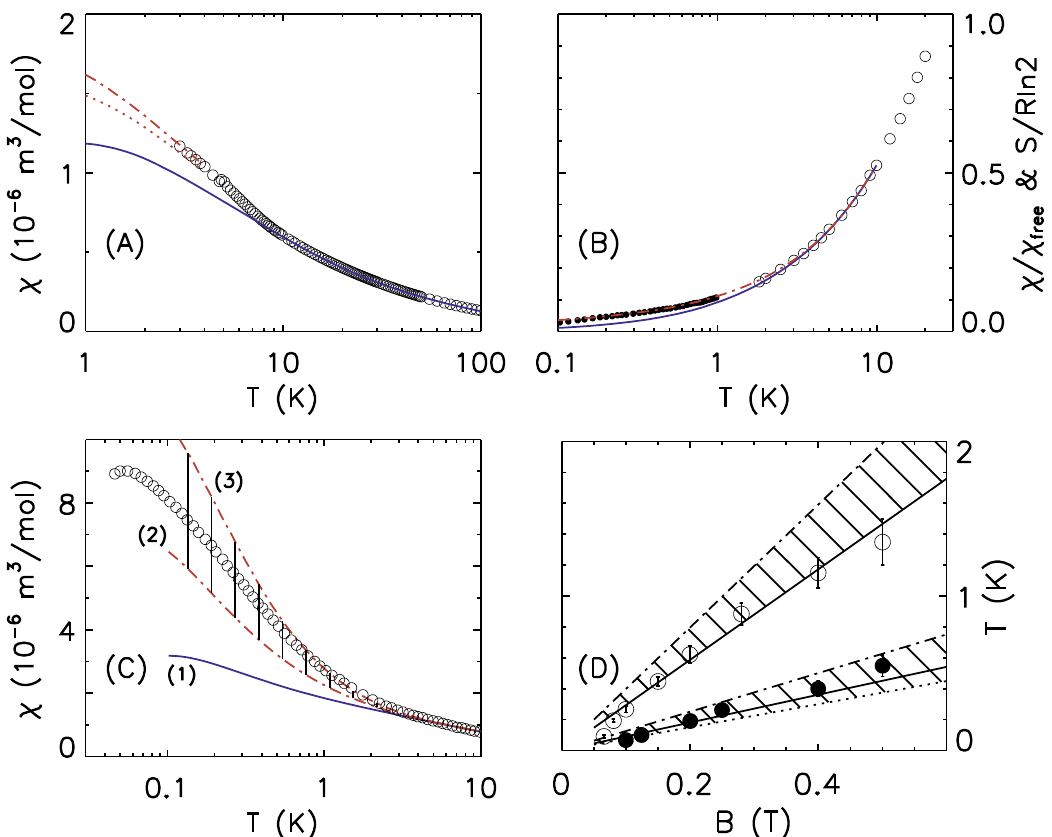}
\end{center}
	\caption{Figure collated and adapted from Ref. \cite{Bretana2023}. Panel (A) displays the measured susceptibility for Ce(Ru$_{0.24}$Fe$_{0.76}$)$_2$Ge$_2$  \cite{Montfrooij2019}, panel (B) for UCu$_4$Pd at $B$= 0.5 T (open symbols)\cite{Bernal1995} and at 0.01 T (filled symbols) \cite{Vollmer2000}, panel (C) for YbRh$_2$Si$_2$\cite{Gegenwart2006} at $B$= 0.025 T.  The solid blue curves are the prediction of eqn. \ref{main} for the contribution of the infinite cluster using the measured entropy $S$ with $\mu_{\text{eff}}$= 1.43, 0.80, and 1.74 $\mu_B$, respectively.  The red dashed-dotted curves in panels (A), (B), and (C) are the response of the superspin collective (evaluated at the percolation threshold) added to eqn. \ref{main}, using the same effective moments. The range in panels (A) and (C) arises from the fact that the simulations were done for an AF-arrangement (see Ref. \onlinecite{Bretana2023} for further details). We opted for a slightly different plotting in panel (B) to highlight the perfect agreement for $T >$ 3 K between the measured entropy and the measured susceptibility (scaled using the susceptibility of a free moment). Panel (D): The field dependence of the maxima in $c/T$ (closed circles)
and ac-susceptibility (open circles) \cite{Custers2010}. The lines denote
the maxima for a simulated distribution of clusters at $p_c$ for a ground
state doublet ($\mu_{\text{eff}}$ = 1.74$\mu_B$), with the hatched areas indicative of
the sensitivity \cite{Bretana2023} of the predicted maxima to the ordering wave vector
(dotted curve: AF ordering; solid curves: replacing the AF-zero
moment of clusters of size 2 with the average moment calculated
for Q = (0.14, 0.14, 0); dashed-dotted curves: replacing clusters
up to size five.}
	\label{ybrhsi}
\end{figure}

In order to make a (quantitative) comparison with the cluster scenario in the absence of direct neutron scattering evidence for the presence of clusters the following workaround was employed \cite{Bretana2023}. From computer simulations the change in entropy as a function of number of unshielded magnetic moments (that is, the occupancy $p$ in percolation language) was calculated by including the shedding of magnetic entropy as soon as an isolated cluster appeared and all its moments aligned as dictated by finite size effects \cite{Fayfar2021a}. This simulated entropy $S(p)$, and derived magnetic specific heat, was then equated to the experimental temperature-dependent specific heat to establish the temperature-dependent occupancy $p(T)$ of the system under comparison. Using this $p(T)$, the simulated ac- and dc-magnetic susceptibility $\chi(p(T))$ could then be directly compared to the measured ones as a function of temperature and applied magnetic field $B$, using the moment $\mu_{\text{eff}}$ of a single unshielded ion as the only adjustable parameter in this comparison. This adjustable parameter was inferred at higher temperatures before isolated clusters with superspins are expected to emerge. At these temperatures, the loss of entropy is due to the shielding of the moments, and the susceptibility reflects the response  $\chi_{\text{free}}$  of the remaining unshielded, uncorrelated moments in the infinite cluster. In detail, for 2-level system with entropy $S$ we have at the higher temperatures \cite{Bretana2023}
\begin{equation}
\chi(B,T)=\frac{S}{R\text{ln2}}\chi_{\text{free}}(B,T,\mu_{\text{eff}}) \approx \frac{S}{R\text{ln2}}\frac{\mu^2_{\text{eff}}}{k_B T}
\label{main}
\end{equation}
At lower temperatures, there is an additional term $\chi_{\text{cluster}}(B,T,\mu_{\text{eff}})$ that reflects the response of the collection of cluster superspins to the external field. We demonstrate this procedure in Fig. \ref{ybrhsi}.

The comparisons produced good quantitative agreement \cite{Bretana2023} for all systems it was applied to: the disordered compounds  Ce(Ru$_{0.24}$Fe$_{0.76}$)$_2$Ge$_2$  and UCu$_4$Pd, and the stoichiometric compounds CeCu$_6$, CeRu$_2$Si$_2$, and YbRh$_2$Si$_2$. The comparisons with Ce(Ru$_{0.24}$Fe$_{0.76}$)$_2$Ge$_2$ (Fig. \ref{ybrhsi}a), UCu$_4$Pd  (Fig. \ref{ybrhsi}b) and CeCu$_6$ \cite{Bretana2023} were fairly straightforward as these systems are at or close to the anti-ferromagnetic order the computer simulations were performed for, the comparison with YbRh$_2$Si$_2$  (Fig. \ref{ybrhsi}c) was slightly more complicated as this system has an incommensurate ordering wave-vector \cite{Stock2012} far removed from the AF-ordering wave-vector, resulting in a band in the comparison (Fig. \ref{ybrhsi}c). We refer the reader to Ref. \onlinecite{Bretana2023} for details. We highlight the comparison for YbRh$_2$Si$_2$ below as it is particularly revealing in how much the cluster scenario simplifies the interpretation of the low-temperature behavior of quantum critical systems.

Figure \ref{ybrhsi}c shows the measured uniform susceptibility compared with the  specific heat measurements for YbRh$_2$Si$_2$ \cite{Custers2003,Gegenwart2005,Custers2010}. At higher temperatures where very few isolated clusters exist the two track each other quantitatively: both curves reflect the response of the remaining unshielded moments of the infinite cluster, a situation also encountered in Ce(Ru$_{0.24}$Fe$_{0.76}$)$_2$Ge$_2$ (Fig. \ref{ybrhsi}a)  and UCu$_4$Pd (Fig. \ref{ybrhsi}b). The scaling factor between the two curves at 10 K yields the value for the unshielded moment $\mu_{\text{eff}}$ of the individual magnetic ions, the one adjustable parameter in the cluster scenario. 

The specific heat curves fully determines $p(T)$. At lower temperatures the measured susceptibility curve and the prediction of eqn. \ref{main} deviate precisely where the formation of finite clusters is expected from the percolation analysis, and the magnitude and shape of the deviation are reproduced for all systems (Fig. \ref{ybrhsi}a-c) by the cluster simulations using the effective moment of the unshielded ions (as determined at higher temperatures). We note that the inferred moment value of 1.74 $\mu_B$ compares quite well \cite{explain1} with the independently measured values of 1.76-1.80$\mu_B$ by means of electron spin resonance \cite{Duque2009} and (1.9 $\pm$ 0.1)$\mu_B$ measured by neutron scattering \cite{Stock2012} (this value has only been measured directly in the YbRh$_2$Si$_2$ system). Thus, the cluster scenario linking the specific heat to the susceptibility has in essence zero adjustable parameters, the only uncertainty (the bands in Fig. \ref{ybrhsi}c) stemming from the internal magnetic structure of the clusters: the simulations were performed assuming simple antiferromagnetic order within each cluster, whereas neutron scattering shows that the actual RKKY ordering in YbRh$_2$Si$_2$ is incommensurate \cite{Stock2012} (see Ref. \onlinecite{Bretana2023} for computational details).

Figure \ref{ybrhsi}d displays the field dependence of the maxima in $c/T$ and in the ac-susceptibility for YbRh$_2$Si$_2$ \cite{Paschen2004,Custers2010}. These two distinct temperature scales and their field evolution are well reproduced \cite{Bretana2023} by averaging the response of the simulated cluster distribution over all superspin sizes, without any further assumptions. Thus, these two energy scales in YbRh$_2$Si$_2$ are not attributable to new physics, the two separate peaks are simply caused by the fact that the specific heat and susceptibility peak at different temperatures in any two level system, given by the possible orientations of the superspin of a cluster in a 2-level system \cite{Stock2012}.

These results demonstrate that the complex phenomenology of quantum critical heavy-fermion systems arises not (necessarily) from novel uniform quantum critical states \cite{Si2024}, but from the unavoidable breakdown of spatial uniformity at the microscopic level. By explicitly including zero-point motion and finite size effects the physics of both chemically substituted as well as stoichiometric systems can be captured in a quantitative self-consistent manner without invoking additional assumptions or ordering mechanisms. Therefore, it appears that the zero-point motion induced local Kondo energy scales play an essential role in the low-temperature response at the expense of the validity of the periodic potential ansatz. In the following section we do a side by side comparison of the uniform and cluster scenarios and the assumptions embedded in them.

\section{Cluster scenario versus uniformity}
The non-Fermi liquid behavior of quantum critical systems has been extensively documented in the literature \cite{Sachdev1999,Stewart2001,Miranda2005,Stewart2006}, together with the new theoretical insights endeavoring to make sense of the new physics \cite{Si2024} emerging at low temperatures. As we discussed in the previous sections, theories pertaining to stoichiometric systems are all based on the assumption of uniformity, an assumption that we have questioned in here and in our previous publications. We now make a detailed comparison of all the salient experimental interpretations for the three competing scenarios. 

We emphasize that we do not question the validity of any of the data collected nor of the quality of the stoichiometric samples: what we do question is the interpretation of the data reflecting an intrinsically uniform system as if all quantum uncertainty associated with the ion positions can be ignored from the outset. We start with a bird's eye view of the competing scenarios.

\subsection{General comparison}

In the Hertz--Millis--Moriya (HMM) framework, quantum criticality arises from the softening of an itinerant magnetic order parameter within an otherwise well-defined Fermi liquid. Starting from a Landau description of the magnetic instability, the fermionic degrees of freedom are integrated out, yielding an effective action for the bosonic order parameter whose dynamics are governed by Landau damping from particle--hole excitations. The resulting dynamical susceptibility for $\mathbf{q}$ near the AFM-ordering wave vector $\mathbf{Q}$ (with $\gamma$ the Landau damping strength) takes the form
\[
\chi^{-1}(\mathbf q,\omega) = \chi^{-1}(\mathbf Q,0)+ c(\mathbf q-\mathbf Q)^2 - i\gamma \omega ,
\]
reflecting overdamped critical fluctuations with dynamical exponent $z$ determined by the damping mechanism ($z$= 2 for AFM). In this picture, the Fermi surface remains intact across the transition, quasiparticles survive away from the ordering wavevector $\mathbf Q$, and non-Fermi-liquid behavior arises from scattering of electrons off critical order-parameter fluctuations. The criticality is thus spatially extended (peaked at $\mathbf Q$) but Gaussian in structure, with interactions entering primarily through self-consistent renormalization of the order-parameter propagator.

In the local quantum critical (LQT) or Kondo-destruction framework, the quantum critical point involves not only the softening of magnetic order but also the collapse of the Kondo screening scale itself. Within extended dynamical mean-field theory (EDMFT), the lattice problem is mapped onto a self-consistent Bose--Fermi Kondo impurity model in which a local moment couples both to a fermionic bath (conduction electrons) and to a critical bosonic bath representing collective spin fluctuations. Inspired by experiments on CeCu$_{5.9}$Au$_{0.1}$ \cite{Schroder2000,Si2001}, the bosonic bath is taken to be momentum independent, so that its critical fluctuations are spatially local but long-ranged in time. This locality is preserved under the EDMFT self-consistency condition, and the anomalous temporal dynamics of the bosonic bath directly determine the impurity spin dynamics. As a result, the local spin susceptibility acquires nontrivial power-law frequency dependence, and exhibits $E/T$ scaling at the interacting fixed point. In contrast to HMM, criticality in this picture is inherently non-Gaussian and involves intertwined spin and hybridization fluctuations, with the vanishing of the Kondo scale leading to a reconstruction of the Fermi surface and the disappearance of quasiparticle coherence at the transition. Thus, within this framework, critical fluctuations of the self-consistent bosonic bath compete with Kondo screening and destabilize the Kondo strong-coupling fixed point, leading to a collapse of the Kondo scale at the quantum phase transition.

In the cluster, percolative framework, quantum criticality arises from the geometric fragmentation of the magnetic lattice rather than from the softening of a uniform order parameter or the collapse of a homogeneous Kondo scale. Spatial variations in the local Kondo scale generate regions in which moments remain unscreened and form correlated clusters embedded in a background of Kondo-shielded sites. The quantum critical point corresponds to the percolation threshold of the infinite cluster, which is intrinsically scale-free and fractal. In contrast to both HMM and LQT, critical behavior is governed both by the details of the breakup of the infinite cluster as well as by the dynamics of isolated clusters. Non-Fermi-liquid behavior emerges from the minting of isolated clusters of increasing size (and associated energy scales), without requiring a uniform collapse of the Kondo scale across the lattice.

\subsection{Detailed comparison}

While the stoichiometric theories are very successful at providing intuitively appealing descriptions of individual phenomena in individual systems, we have argued that there exists a simpler description based on the chaotic environment experienced by the conduction electrons, characterized by the fragmentation of the lattice into isolated clusters. Note that this is fundamentally different from Griffiths phase physics \cite{Miranda2005,CastroNeto2000} as in the cluster scenario the diverging response is not due to rare regions. However, the cluster scenario also requires additional assumptions beyond the existence of a fleeting distribution of Kondo energy scales in order to describe all of the data. We summarize the interpretations of experimental data according to uniform scenarios and the cluster scenario in Table \ref{comparison} and expand on the details below, emphasizing the interpretations in the cluster scenario as those are probably the most controversial.

\onecolumngrid

\begin{table*}[t]
\caption{Interpretations of data within opposing scenarios. HMM is short for Hertz-Millis-Moriya scenario \cite{Hertz1976,Millis1993,Moriya1985}; LQT for Local Quantum Criticality \cite{Si2001,Paschen2004,Si2024}; FS for Fermi surface and QCP for Quantum Critical Point.}
\label{comparison}
\small
\begin{tabular}{p{0.31\textwidth} p{0.31\textwidth} p{0.31\textwidth}}
\toprule
\textbf{Experimental observation} & \textbf{Interpretation assuming uniformity} & \textbf{Interpretation with Kondo distribution} \cite{Heitmann2014,Fayfar2021a,Bretana2023} \\
\midrule
Chemically doped and stoichiometric systems display same phenomenology & Behaviors assumed to be intrinsically different & Phenomenology independent of whether Kondo distribution is static or dynamic \\
Number of correlated moments independent of crystallographic direction & To our knowledge not addressed explicitly & Moments on randomly formed clusters are aligned due to finite-size effects \\
Dynamics shows both short-lived and very long-lived excitations & To our knowledge not addressed explicitly & Depending on size of superspin, it fluctuates or is frozen out \\
Divergence of $c/T$ & Bosonic critical fluctuations (HMM), critical collapse of Kondo scale and associated divergence of quasiparticle mass (LQT) & Breakup of infinite, disordered cluster into finite, ordered clusters of increasing size \\
Coherence temperature in resistivity & Onset of coherent shielding of moments & Coherent shielding and moments on isolated clusters align, reducing scattering mechanism \\
Low-$T$ resistivity deviates from $T^2$, down to linear & Electrons scatter off critical fluctuations (HMM); breakdown of quasiparticles, entire FS critical (LQT) & Resistivity has a component determined by number of moments on infinite cluster \\
Rapid changes in Hall coefficient & Strong changes in scattering at FS hot spots (HMM); Kondo breakdown, reconstruction of FS (LQT) & Non-analytic disintegration of infinite cluster; field-induced polarization of superspins \\
Magnetic moments inferred from different techniques do not match & Curie-like moments follow from critical magnetic fluctuations, saturation moments reflect lack of long-range order & Follows from lattice fragmentation into isolated clusters, uncorrelated unshielded moments, and shielded moments \\
Field-dependent ac-susceptibility peaks at temperatures too large to be associated with individual moments & Reflects crossover scale between quantum critical and Fermi liquid regimes (HMM); between heavy Fermi liquid and local moment regimes (LQT) & Reaction of superspins of clusters to field \\
Field-dependent $c/T$ and ac-susceptibility do not peak at same temperature & Maxima relate to different derivatives of free-energy scaling function, ratio of slopes not universal & A collection of superspins peaks at different temperatures, ratio of slopes fixed by distribution and ordering wavevector \\
System driven to order sheds very little entropy & Kondo screening has already quenched most of local-moment entropy & Kondo screening and cluster formation have removed most of the entropy; only superspin orientational entropy left \\
Hyperscaling in $E/T$ and $B/T$ & Quasi-2D or crossover regions mimicking scaling (HMM), Kondo scale collapse at QCP and emergence of interacting fixed point (LQT) & Superspin reorientations provide the zero-energy degrees of freedom required for high-energy-like response \\
\bottomrule
\end{tabular}
\end{table*}

\twocolumngrid

\textbf{Similarity of behavior of doped and stoichiometric compounds.} Experiments on chemically doped quantum critical systems and on stoichiometric systems driven to quantum criticality by applying hydrostatic pressure revealed very similar behavior. For instance, S\"{u}llow \textit{et al.} \cite{Sullow1999} showed that the phasediagram obtained by applying pressure to CeRu$_2$Ge$_2$ was indentical to the phase diagram obtained by applying chemical pressure through iso-electronic substitution of Ge atoms by smaller Si atoms in CeRu$_2$(Ge$_{1-x}$Si$_x$)$_2$. Neutron scattering results also revealed the close connection between doped and stoichiometric systems \cite{Bretana2021}. Both disordered \cite{Aronson1995,Montfrooij2003} and stoichiometric systems \cite{Schroder2000} can display hyperscaling. However, doped and stoichiometric systems have been treated differently when it comes to theoretical descriptions. For example, even though the chemically disordered compound UCu$_4$Pd was successfully described \cite{Bernal1995} by invoking a distribution of Kondo energy scales associated with chemical disorder, there have been no attempts to try to describe experimental data on stoichiometric compounds in terms of Kondo distributions (besides our own attempt \cite{Bretana2023}). As we argued in the preceding, a distribution of Kondo energy scales unavoidably materializes and by extension, (some of) the behavior associated with disorder should have its reflection in the data. The cluster scenario offers a natural explanation for why chemically disordered systems behave so similar to stoichiometric compounds since to the fast moving electrons there is no difference between static disorder and picosecond-scale dynamical disorder. In contrast, assuming uniformity from the outset does not explain why disordered systems react so similarly to experimental probes. Any successful theory based on uniformity should also be able to describe why non-uniform systems behave alike.

\textbf{Independence of correlation lengths on interaction strength.} 
Neutron scattering data on the chemically doped tetragonal compound Ce(Ru$_{0.24}$Fe$_{0.76}$)$_2$Ge$_2$ \cite{Montfrooij2007} as well as on the  stoichiometric system CeRu$_2$Si$_2$  \cite{Kadowaki2006,Bretana2021} revealed the onset of magnetic correlations at incommensurate ordering vectors. This onset took place below the coherence temperature and its intensity continued to increase with decreasing temperature. Remarkably, the number of moments that were correlated did not display any directional dependence \cite{Bretana2021}, even though the strength of the RKKY ordering interaction \cite{Ruderman1954,Kasuya1956,Yosida1957,Endstra1993}  depends on the crystallographic direction. This equality was observed to hold for any temperature where the correlations could be observed \cite{Montfrooij2019}. This appearance of cubic symmetry in tetragonal systems, together with the fact that the correlations remain short-ranged is as expected for the cluster scenario in which clusters form according to random patterns with the moments on isolated clusters aligning  and forming superspins. As far as we are aware, these findings are not discussed in uniform scenarios. Unfortunately, there do not appear to exist more neutron scattering datasets where correlation lengths have been measured along different directions in non-cubic systems.

\textbf{Coexistence of static and dynamic short range order.}
Depending on the magnitude of any particular superspin, it is free to fluctuate or it will be frozen, even at absolute zero as discussed by Hoyos and Vojta \cite{Hoyos2006}. Experimental evidence supporting the coexistence of a small fraction of frozen superspins with non-frozen superspins comes from neutron scattering experiments on non-stoichiometric Ce(Ru$_{0.24}$Fe$_{0.76}$)$_2$Ge$_2$ where it was observed \cite{Heitmann2016,Montfrooij2019} that the clusters that form at the lowest temperatures do not fluctuate on the time scale of the neutron scattering experiment (80 ps) while the ones that had already formed at higher temperatures continue to fluctuate. This is interpreted in the cluster scenario as the clusters that form at the lowest temperatures are the ones that are most likely to be frozen \cite{Hoyos2006} as they are the ones with most members $s$, and therefore, the ones with the largest superspin as the superspin roughly depends on the cluster size as $\sqrt{s}$. However, we note that -- to our knowledge -- equivalent experiments on stoichiometric systems do not  exist.

\textbf{Divergence of $c/T$ at the quantum critical point.} 
Heavy-fermion systems are named after the large value of the $c/T$ coefficient that reaches values hundreds of times more \cite{Stewart2001} than expected for metals described by Fermi-liquid theory \cite{Landau1957}, and experimental data are consistent with this coefficient diverging at the quantum critical point. The divergence has the most natural explanation in uniform scenarios as it reflects what critical modes are expected to do near a phase transition, independent of the exact character of the critical modes. The enhanced value of the $c/T$ coefficient and interpretation of electrons acquiring a large effective mass is interpreted as either the signs of critical fluctuations in bosonic modes \cite{Millis1993} or the breakdown of quasiparticles and the transition to local criticality \cite{Si2024}, an interpretation that is intuitively appealing as heavy particles move more slowly, and effectively become localized when $c/T$ diverges.

The interpretation of the same data in the cluster scenario is vastly different: the enhanced $c/T$ coefficient and its divergence are determined by the critical behavior of the infinite cluster near the percolation threshold. Entropy is released when a magnetic moment on the infinite cluster is being shielded according to the local Kondo energy scale, while the rapid increase in entropy release (which results in the large value of the $c/T$ coefficient) is due to the moments of newly formed isolated clusters lining up with their neighbors \cite{Fayfar2021a}. Thus, what is interpreted as a large effective mass of the Bloch electrons is instead associated with the magnetic ordering of the moments on newly minted clusters in the cluster scenario. In addition, the divergence of the number of moments being locked up in isolated clusters close to the percolation threshold is responsible for the divergence of the $c/T$ coefficient. Provided the Kondo distribution can be expanded in power of $T$ (something which is not possible at $T$= 0) $c/T$ would exhibit a power law dependence at the percolation threshold: $c/T \sim |T-T_c|^{\beta_T-1}$ \cite{footnote}.

\textbf{Disparate moments.}
The magnetic response is a combination of that of the individual moments on the infinite cluster as well as the superspins of the isolated clusters. Depending on whether we determine the effective magnetic moment from a Curie-type of law, saturation magnetization, or from the peak in the ac-susceptibility, we obtain seemingly different moment values by many orders of magnitude. However, when we account for the various fractions of the moments that are shielded, free to fluctuate, or locked up in isolated clusters, then  the inferred values converge onto a single one, as demonstrated \cite{Bretana2021} in CeRu$_2$Si$_2$ \cite{Takahashi2003}.   Also, the existence of the two components in the magnetic response could be clearly identified in YbRh$_2$Si$_2$ as a function of magnetic field (See Fig. 5 in Ref. \onlinecite{Bretana2023}) by applying fields large enough to freeze the superspins at a given temperature. 

The stoichiometric compounds CeRu$_2$Si$_2$ \cite{Takahashi2003},  YbRh$_2$Si$_2$ \cite{Gegenwart2005}, and CeCu$_6$  \cite{Tsujii2000} all display maxima in the field-temperature dependent ac-susceptibilities that correspond to moment values exceeding \cite{Bretana2023} the average rare-earth moment. In uniform scenarios this scale is taken as an crossover energy scale pertaining to the crossover between the Fermi-liquid and quantum critical regimes (HMM) or between a small and a large Fermi-surface (LQT). In the cluster scenario, the ac-susceptibilty peaks at the average superspin moment, which simulations showed \cite{Bretana2021,Bretana2023} to exceed the individual moments even in an AF-system. We note (again) that this predicted peak position is fully determined by the size of an individual fluctuating moment, it does not represent an (additional) adjustable parameter. As such, the existence of these 'supersized' moments and their numerical values serves as a strong check on the cluster scenario.

\textbf{Superparamagnetic/ferromagnetic fluctuations.}
In  YbRh$_2$Si$_2$ it was measured \cite{Gegenwart2005} that the magnetic field dependent Wilson ratio was in fact field-independent for fields $B > 0.2$T at low temperatures, but showed strong enhancements at lower fields.  At the same time, the ratio A/$\chi_0^2$ (with A the coefficient of the $T^2$ term in the resistivity) was approximately constant. This was interpreted \cite{Gegenwart2005} as evidence for ferromagnetic $q=0$ fluctuations being strong in a wide region of the phase diagram, while they must  (by necessity) become subleading to $q=Q$ fluctuations asymptotically close to the QCP. We refer to Ref. \onlinecite{Gegenwart2008} for more details, as well as to how this observation might be placed within the HMM scenario.

In the cluster scenario, a similar behavior is expected when at high $B/T$ ratios the superspins are frozen (aligned with the field), whereas at low fields they can react to the field in a paramagnetic manner. Using the results of percolation simulations, Breta\~{n}a et al. showed \cite{Bretana2023} that the field dependence of the Wilson ratio displayed numerical agreement with the simulated collection of superspins (see Fig. 5 in Ref. \cite{Bretana2023}). They also showed agreement between this scenario and the ac-susceptibility results as a function of field and temperature for the stoichiometric systems  CeRu$_2$Si$_2$ \cite{Takahashi2003}and CeCu$_6$ \cite{Tsujii2000}. Therefore, the cluster scenario offers a quantitative interpretation of the divergence of the Wilson ratio without invoking any additional (subleading) $q$= 0 fluctuations.

We note that the interpretation of ferromagnetic fluctuations appears to run contrary to the ac-susceptibility results, at least following a straightforward albeit heuristic argument. In YbRh$_2$Si$_2$ (as well as in  CeRu$_2$Si$_2$ and CeCu$_6$) the low temperature ac-susceptibility curves peak (for fixed frequency) at increasingly lower temperatures with \textit{decreasing} fields. If YbRh$_2$Si$_2$ truly harbored ferromagnetic fluctuations rather than superspins associated with a distribution of clusters, then we would expect that the peak positions would shift to lower temperatures with \textit{increasing} fields as the correlation length $\xi$ of the critical fluctuations would be increasingly limited by the increasing magnetic field. This limitation ensures that the relaxation time $\tau(T)\sim \xi^z$ of a correlated volume will not get as slow as it would without a field, and therefore, the peak position of $\chi_0(T)/(1+\omega^2\tau(T)^2)$ (with $\omega$ the ac-frequency) would occur at a lower temperature with field than without field, provided $\chi_0(T)$ is diverging.

\textbf{Multiple energy scales.}
The existence of two different energy scales for the ($B$,$T$)-peak positions in the $c/T$ and ac-susceptibility curves (see Fig. \ref{ybrhsi}d) is not only consistent with the superspin-cluster scenario, the scenario would have been refuted had these different scales not been observed. For instance, for any two level system, the specific heat $c/T$ and the ac-susceptibility peak at $\mu\mu_B B=1.622$k$_BT$ and  $\mu\mu_B B=0.772$k$_BT$, respectively, with $\mu$ the magnetic moment. For a collection of superspins the ratio between the two peak positions will change, but they cannot coincide. We showed the response (and good agreement) for YbRh$_2$Si$_2$ compared to a collection of superspins at the percolation threshold in Fig. \ref{ybrhsi}d.

The data for the maxima of the ac-susceptibility $T^*(B)$ as shown in Fig. \ref{ybrhsi}d appear to follow a linear relation  $T^*(B) \sim |B-B_c|$ with $B_c$ the critical field \cite{Custers2010}. This linear relationship is in agreement with all three scenarios, but for entirely different reasons. In the cluster scenario, the peak is caused by Zeeman splitting directly leading to a linear relation as discussed in the previous paragraph. In the HMM scenario, the peak position is expected \cite{Millis1993,Sachdev1999} to be given by $T^*(B) \sim |B-B_c|^{\nu z}$,  with the dynamical exponent $z$= 2 in an itinerant AFM metal and the mean field correlation-length exponent $\nu$= 1/2 for the Gaussian fixed point, leading to a linear behavior $\nu z$= 1. In the LQT scenario, we would not expect $\nu z$ to be equal to 1, but since temperature is the only energy scale \cite{Si2024} resulting in B/T-scaling, the linear behavior would be a manifestation of this scaling. 

\textbf{Loss of entropy at ordering.}
Systems near the quantum critical point can be driven into an ordered phase by judiciously changing external parameters such as magnetic field, temperature, or hydrostatic pressure.  When this was done in YbRh$_2$Si$_2$, it was measured that the amount of entropy released was exceptionally small at $\sim$0.008 Rln2 \cite{Custers2003}.  In the cluster scenario, such a small number is to be expected as the only entropy remaining in the system at the percolation threshold is that of the reorientation degree of freedom of the superspins of the isolated clusters. According to a percolation simulation of a 2-level system on a body-centered lattice \cite{Fayfar2021a}, this remaining entropy at the percolation threshold equals 0.0078 Rln2. Thus, in the cluster scenario we can have a very small number on the one hand (entropy), as well as a very large number on the other hand (effective moments exceeding the bare rare-earth moment as obtained from ac-susceptibility measurements) without needing to invoke any further assumptions.

\textbf{Resistivity and Hall coefficient.}
Resistivity and Hall effect measurements are not nearly as clear to interpret in the cluster scenario as the measurements discussed  above, let alone confirm it. The multiple components of the cluster scenario (infinite cluster, shielded moments, frozen and unfrozen superspins) are all expected to react differently to these probes. Any sharp features \cite{Paschen2004} would have to be ascribed to geometric criticality when the infinite cluster is about to break up or to the ordering of the superspin moments, and unusual exponents of the longitudinal resistivity can find their origin in the temperature dependence of the infinite cluster. For example, in the temperature range where $c/T$ is almost constant with an enhanced ratio representing a heavy-fermion system, this would imply that the entropy, and therefore the strength of the infinite cluster, varies linearly with temperature. Thereby, the component of the resistivity linked to the scattering rate by individual, fluctuating moments on the infinite cluster also acquires an underlying linear temperature dependence. 

One of the most revealing experiments on quantum critical systems are the Hall probe experiments on YbRh$_2$Si$_2$  \cite{Paschen2004} as a function of temperature and field. These experiments showed that the Hall coefficient is marked by two distinct regions in the ($B,T$)-phase diagram, with a crossover range between the two that becomes sharper in field with lower temperature, encompassing the possibility that there exists a zero-width change in slope at the QCP. Interpreting the slope with field as a measure of the number of charge carriers, this implies a sudden reconstruction of the Fermi surface at the QCP \cite{Paschen2004}. These result feature prominently \cite{Si2024} in the LQT scenario.

The cluster scenario should also be accompanied by a marked variation in the Hall effect. A heavy-fermion system in the presence of isolated clusters with superspins, an infinite cluster, and shielded moments should display both the normal and the anomalous Hall effect (see methods in Ref. \onlinecite{Paschen2004}), with the anomalous effect determined by the remaining moments on the infinite cluster as well as by the superspins. The moments on the infinite cluster are in the process of being Kondo shielded, but when some of them are spun off into isolated clusters, their moments become protected from Kondo shielding while the carrier density goes down. Thus, we can expect the field behavior of the Hall coefficient to be dependent on temperature, even strongly so.

To understand the field dependence of the Hall effect regarding the observed crossover \cite{Paschen2004}, we focus on the collection of superspins at very low temperature. Here, as the applied magnetic field is increased, progressive polarization of clusters suppresses both magnetization susceptibility and longitudinal resistivity, leading to a crossover in the Hall slope as a function of applied field. At high fields, once superspins are saturated, the anomalous contribution becomes nearly constant and the Hall signal is dominated by the ordinary term.

The field scale associated with the polarization of any given cluster moment $\mu$ is set by the competition $\mu B \sim k_B T$, so that the characteristic crossover field obeys $B^*(T) \sim k_B T/\mu$. The superspin distribution is dominated by the rapidly fluctuating smaller clusters (see Fig. 8 in Ref. \onlinecite{Bretana2021}), whereas the largest clusters that form closest to the percolation threshold are most likely to be frozen out \cite{Hoyos2006}. Therefore, near $p_c$, the collection of superspins that is allowed to fluctuate will not be strongly temperature dependent so that the averaged polarization crossover field $B^*$  is expected to scale proportionally with temperature. Consequently, the width of the Hall crossover narrows linearly as $T$ decreases, reflecting the reduced thermal energy required to polarize the same (fluctuating) superspin ensemble. This sharpening arises naturally from the Zeeman polarization of cluster moments and does not require \cite{Si2024} a discontinuous reconstruction of the Fermi surface. Note that this crossover scale is essentially the same scale as the ($B$,$T$)-peak scale obtained from ac-susceptibility measurements. 

\begin{figure}
\begin{center} 
\includegraphics*[viewport=40 105 280 300,width=85mm,clip]{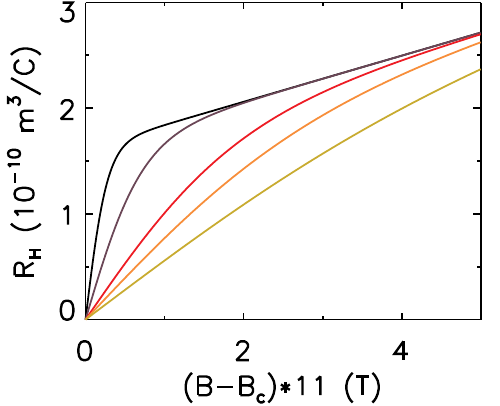}
\end{center}
	\caption{The temperature dependence of the Hall coefficient $R_H$ calculated for a system where a fraction $f$ of 28\% of the local moments remain unshielded at the percolation threshold. The lines have been calculated for T=[29 mK, 75 mK, 0.2 K, 0.3 K, 0.5K] (from left to right), using the same $\mu_{\text{eff}}$ as in Fig. \ref{ybrhsi}. For a 2-level system, $R_H$ is given by $R_H=R_0(1-f)B+a_0 f\mu_{\text{eff}}$tanh$(\mu_{\text{eff}} \mu_B B/k_BT)$, where we have used the $R_0$ value from Ref. \cite{Paschen2004} and taken $a_0=R_0$ for lack of a robust estimate.}
	\label{paschenplot}
\end{figure}

In order to quantify this reasoning, we plot an estimate for the Hall coefficient $R_H(B,T)$ in Fig. \ref{paschenplot}. In the spirit of Ref. \onlinecite{Paschen2004} we write the Hall coefficient as a sum of the ordinary Hall coefficient $R_0$ and a magnetization dependent part, where we have taken the known experession for the field and temperature dependent magnetization of a 2-level system. At the percolation threshold in a body centered system, more than 20\% of the moments remain unshielded, resulting in a marked change of slope between low and high fields.  Note that a fluctuating cluster of $s$ members, independent of whether the moments are aligned or not, represents $s$ fluctuating moments of size $\mu_{\text{eff}}$ that give rise to an anomalous Hall term. In order to facilitate the comparison with Fig. 2a in Paschen \textit{et al.}, we simply took $\mu=\mu_{\text{eff}}$ and we have multiplied the magnetic field by 11 in the plot (not in the calculations) and labeled the axis $B-B_c$. We did not attempt to include the term $R_{\infty}(T)$ \cite{Paschen2004} even though this term represents the part of the temperature dependence that makes $R_H$ increase rapidly with increased temperature; in the spirit of Ref. \onlinecite{Paschen2004} we focus on the change in slope. Therefore, only assuming that a significant fraction of the moments survives down to $T$= 0 is sufficient to reproduce the characteristic change in slope first measured by Paschen \textit{et al.}. We note that including the simulated cluster distribution would result in a smearing of the curves in Fig. \ref{paschenplot}, however, the sharpening trend would not be affected. 

Of course, the Hall data reported in Ref. \onlinecite{Paschen2004} had been corrected for the anomalous Hall effect, using a standard approach based on the measured magnetization (see methods in Ref. \onlinecite{Paschen2004}).  As such, our above discussion and Fig. \ref{paschenplot} might appear to be moot, however, as we showed in Fig. \ref{tokiwaplot}, the magnetization in the critical region is actually given by the superspins of the clusters whose individual moments are fully aligned. Therefore, using the measured magnetization for eliminating the anomalous Hall term underestimates the actual number of fluctuating moments by up to a factor of $\sim$20 as outlined in the discussion around Fig. \ref{tokiwaplot}. Paschen \textit{et al.} estimated the anomalous term to not exceed 20\% of the normal term, but taking this factor into account it is likely that a substantial superspin contribution survives their correction procedure. In all, we conclude that the cluster scenario appears to be consistent with the experiments, even though it fails to provide any quantitative predictions other than potentially the Hall crossover scale. 

Lastly, there is one experimental complication that shows up more readily in resistivity and Hall effect measurements rather than in specific heat and magnetization experiments. The Kondo distribution is dynamic: its shape is determined by the zero-point motion amplitude, but exactly which sites are more prone to Kondo shielding than others varies on the time scales of ionic motion. Whereas this time scale is much slower than that of the motion of an individual electron or a magnetic probe, in resistivity and Hall measurements the time scale of the experiment is such that it actually probes an averaged response, further complicating the interpretation of resistivity measurements and the Hall effect. 

\textbf{Origin of hyperscaling in E/T and B/T.} In some quantum critical systems it was discovered \cite{Aronson1995} that the repsonse of the system when probed by means of inelastic neutron scattering by transferring an amount of energy $E$ to the system at a temperature $T$ was only dependent on the ratio of $E/T$ rather than both variables separately. Field dependent measurements on CeCU$_{5.9}$Au$_{0.1}$ showed \cite{Schroder2000} that a similar scaling existed in $B/T$. These findings implied that temperature was the only relevant variable, the equivalent of high-energy behavior when the energy scales at which the system is being probed greatly exceeds the energy scales associated with the internal degrees of freedom. The cluster scenario offers a fairly straightforward explanation for this, provided that the moments on isolated clusters order.

When the moments on an isolated cluster align, there is still one degree of freedom left: the superspin of the cluster. Reorientations of this superspin in the absence of a magnetic field do not cost any energy \cite{Broholm2002}, thereby providing the internal low-energy degrees of freedom required for high-energy like behavior. Interestingly, this behavior does not appear to be restricted to quantum critical systems as a similar scaling was observed in a classical system characterized by a distribution of magnetic clusters \cite{Schimmel2001,Heitmann2010}.  Also, the partition function of a collection of superspins and individual moments is a function of $B/T$ only, and therefore, when probed using a magnetic field we can expect the superspin response to vary with $B/T$, resulting in (approximate) $B/T$ scaling.

The perspective described in this section suggests that the primary difficulty in interpreting quantum critical heavy-fermion data may not lie in the complexity of the underlying interactions or in Fermi surface topology or instability, but in the restrictive assumption of uniformity imposed at the outset. At the same time, the uniform approach and ensuing Fermi-liquid theory are one of the most successful and experimentally verified aspects of solid state theory. In the remainder of the paper we discuss when exactly we should expect a breakdown of the uniform approach, and whether the organizing principles we use to describe low-temperature behavior survive in the presence of intrinsic quantum disorder.

\section{The periodic potential ansatz and organizing principles}
As discussed, the assumption of a perfectly periodic electronic potential, $V(\mathbf r) = V(\mathbf r + \mathbf R)$, is generically invalid in quantum critical heavy-fermion systems because of the unavoidable quantum-mechanical breakdown of spatial uniformity. Yet this assumption underlies much of modern solid-state theory \cite{AshcroftMermin} and has been extraordinarily successful in describing a wide range of materials. It therefore seems prudent to understand not only where the assumption fails, but also why (and when) it works as well as it does.

We emphasize that we do not claim a loss of translational invariance of the full electron--ion Hamiltonian. Rather, once screened and unshielded moments are spatially intermixed in a non-self-averaging way, the usual Bloch-band quasiparticle construction based on a single periodic effective potential (and a single $T_K$) is no longer a controlled low-energy starting point.

The periodic potential ansatz permits the decomposition of electronic eigenstates into Bloch waves \cite{Bloch1929}. Under this assumption, the single-particle Schr\"odinger equation admits solutions of the form \cite{AshcroftMermin}
\begin{equation}
\psi_{n\mathbf k}(\mathbf r) = e^{i\mathbf k\cdot \mathbf r} u_{n\mathbf k}(\mathbf r),
\end{equation}
where $u_{n\mathbf k}(\mathbf r)$ has the periodicity of the lattice. This result is the mathematical foundation of band theory: it allows the entire electronic structure of a macroscopic crystal to be represented within a single primitive unit cell, and it produces well-defined energy bands, Fermi surfaces, and quasiparticles.  Without the periodic potential assumption, none of this standard machinery of condensed-matter physics can be constructed in any rigorous way.  At the same time, the remarkable empirical success of band theory across metals, semiconductors, and insulators reflects the fact that for a wide class of materials the deviations from perfect periodicity are sufficiently weak or sufficiently short-ranged that Bloch's theorem remains an excellent approximation. This even holds for Kondo systems at temperatures above the average Kondo scale, or for that matter for systems well below the average Kondo scale such as Kondo insulators \cite{Dzero2016}.

The periodic potential ansatz is perhaps best viewed as a controlled approximation whose domain of validity is set by the competition between two length scales: the scale over which the electronic environment varies, and the scale over which the electronic wave functions extend. When electronic states sample many unit cells and the local deviations from periodicity fluctuate on shorter length scales, the electrons effectively self-average over the underlying inhomogeneity. In this regime, the periodic potential approximation becomes asymptotically exact, and the familiar concepts of quasiparticles and bands emerge naturally.

In quantum critical heavy-fermion systems, however, the intrinsic nonuniformity generated by the broad distribution of Kondo temperatures introduces spatial variations on all length scales. Under these conditions, self-averaging fails. The electrons no longer experience a well-defined periodic background, and the conceptual foundations of uniform band structure and Fermi-liquid theory cease to apply in a strict sense.

\subsection{The periodic potential in modern theory}

Landau Fermi-liquid theory presupposes the existence of long-lived quasiparticles that are adiabatically connected to the eigenstates of a noninteracting electron gas in a periodic potential \cite{Landau1957}. The quasiparticle momentum $\mathbf k$ is a good quantum number, the Fermi surface is a sharply defined object in reciprocal space, and the low-energy excitations are labeled by deviations from this surface. All of these constructions rely on translational invariance and the existence of Bloch states. 

Even when interactions are strong, Fermi-liquid theory assumes that the electronic environment remains spatially uniform on the scales relevant for quasiparticle formation. However, once the underlying periodicity is lost or becomes ill-defined, the notion of a globally coherent Fermi surface and a uniform goundstate becomes the wrong starting point for a theoretical description.

The standard theoretical description of heavy-fermion systems begins with the Kondo lattice model and proceeds via mean-field decouplings that produce a spatially uniform hybridization between localized $f$-electrons and conduction electrons. Extensions beyond static mean-field theory, such as dynamical mean-field theory (DMFT)\cite{Georges1996}, retain this assumption of translational invariance at the lattice level, mapping the correlated lattice problem onto a self-consistent impurity model that is identical on every site. This procedure yields heavy quasiparticle bands with well-defined dispersions and Fermi surfaces, again built upon the assumption of translational invariance.

While fluctuations around the mean field are often included, the mean-field solution itself enforces uniformity at the outset. The existence of a single Kondo temperature and a homogeneous heavy-fermion ground state are not results of the theory, but assumptions built into its construction.

Density functional theory (DFT) inherits the periodic potential assumption at its most basic level. In practical implementations for crystalline solids \cite{Ullrich2012,ShollSteckel2022}, one solves the Kohn--Sham equations \cite{KohnSham1965} using basis functions that are explicitly Bloch-periodic, and the electronic density is constrained to obey the same translational symmetry. Exchange--correlation functionals are constructed under the implicit assumption that the system can be locally approximated by a homogeneous electron gas, an approximation that becomes increasingly accurate when the electronic environment varies slowly on atomic length scales.

Thus, DFT does not merely exploit periodicity as a computational convenience; it is built upon it as a structural assumption. The extraordinary success of DFT in predicting equilibrium structures, phonon spectra, elastic constants, and electronic band structures again reflects the fact that for many materials the ground state is sufficiently close to uniform on microscopic scales.

The failure of the periodic potential ansatz in quantum critical heavy-fermion systems does not imply that existing theoretical frameworks should be discarded. Rather, it calls for identifying organizing principles that remain valid when strict translational invariance is no longer a suitable starting point. Several such principles have emerged in the literature; viewed from the perspective developed here, they can be interpreted as partial and phenomenological attempts to repair the breakdown of uniformity. 

\subsection{Limitations of Hertz--Millis quantum criticality}

The Hertz-Millis-Moriya framework \cite{Hertz1976,Millis1993,Moriya1985} assumes that the low-energy physics near a quantum critical point is governed by spatially extended, weakly damped collective fluctuations of a single itinerant electronic fluid. This concept presupposes a uniform underlying Hilbert space in which all electronic degrees of freedom participate coherently in the critical dynamics.

The present results demonstrate that this assumption fails in quantum critical heavy-fermion systems. As the temperature is lowered, the Hilbert space fractures into qualitatively distinct sectors: individual moments that are Kondo shielded, superspins associated with finite clusters, and fluctuating moments residing on the lattice-spanning infinite cluster. Even at the percolation threshold, the dynamics are not uniform. Depending on the size of the superspin, some superspins will be frozen \cite{Hoyos2006} on the time scale of any experiment, while others are free to reorient. The frozen sector contributes static degrees of freedom, while the fluctuating sector partakes in the low-energy dynamics. Critical behavior therefore emerges from the evolving connectivity between these sectors, rather than from uniform order-parameter fluctuations of a single fluid.

In such a setting, the basic premise of the HMM construction — that the system can be described by a homogeneous effective field theory for a collective order parameter — no longer applies. The failure is not quantitative but structural, and it originates from the intrinsic breakdown of spatial uniformity at the microscopic level.

We note that critical behavior in the cluster scenario is unwieldy. First, we have the breakdown of the infinte cluster which comes with its own critical exponents \cite{StaufferAharony,Fayfar2021a,Fayfar2021b}. This type of breakdown has been shown to be in a universality class that does not satisfy the Harris criterion \cite{Harris1974} and as such, exponents are system dependent rather than universal. Second, the dynamics of the superspins of the emerging isolated clusters are themselves critical \cite{Hoyos2006}. As such, it is unlikely that geometric criticality can be easily separated from the concurrent superspin critical behavior.

\subsection{Local quantum criticality and quasiparticle breakdown}

Local quantum criticality emphasizes the failure \cite{Si2001,Paschen2004,Si2024} of spatially extended order-parameter descriptions and the emergence of critical dynamics that are local in space but critical in time.  Such a scenario relaxes the spatial extension of order-parameter fluctuations but not the spatial uniformity of the underlying electronic environment. The collective magnetic fluctuations couple to the local moments that are in the process of being Kondo screened. Thus, the local moments sit in a bosonic bath that impedes Kondo screening. As a result when these fluctuations become sufficiently strong (near the QCP), the Kondo energy scale collapses and the local spin susceptibility acquires a singular power-law form. The singular local susceptibility generated by Kondo destruction feeds back through the EDMFT self-consistency condition that relates the local susceptibility to the momentum-dependent lattice susceptibility. The resulting bosonic propagator thereby also becomes locally critical. The concept of Kondo breakdown implies that the heavy quasiparticles themselves disintegrate at the quantum critical point \cite{Si2001,Si2024}, leading to a reconstruction of the Fermi surface. In all, LQT is a very elegant mechanism through which extended magnetic fluctuations in a perfectly periodic structure can produce locally critical degrees of freedom. 

This viewpoint has some overlap with an intrinsically nonuniform ground state composed of clusters and superspins. In such a system, critical fluctuations are governed not by a uniform order parameter, but by the collective dynamics of spatially disconnected clusters whose remaining degrees of freedom dominate the low-energy response. However, there is an important difference between the cluster scenario and the local critical scenarios: in the cluster scenario, local effects are driven by  the inherent quantum chaotic nature of the ground state (zero-point motion) and they change both randomly from place to place, and from time to time, whereas in the uniform scenario they are uniform in space. In addition, this underlying driver of local behavior, the ionic zero-point-motion that changes the instantaneous landscape of the cluster distribution is in itself not critical as it occurs on a much slower time scale than that of the electronic motion. Within the cluster framework, there exists a more natural interpretation: the apparent breakdown reflects the progressive loss of the lattice-spanning cluster as the system is tuned through its percolative transition. Rather than representing a uniform collapse of hybridization, the observed signatures of Kondo breakdown arise from the geometric disintegration of the coherent electronic network combined with an emerging population of superspins. In reality, about three-quarters of the system remains Kondo shielded at the percolation threshold, while the remainder of the moments is protected from Kondo shielding by being in an ordered environment. But note that there has been no change in any of the physical interactions that determine whether moments are being Kondo screened or not, or in the nature of any quasiparticle.

\subsection{Ferromagnetic fluctuations in an anti-ferromagnetic system}

The emergence of superspins for isolated clusters naturally results in a super-paramagnetic response for a quantum critical system when probed in the presence of a magnetic field. For an anti-ferromagnetic system, a cluster of size $s$ will have dangling (uncompensated) moments, giving the cluster a superspin that to good approximation scales as $\sim\sqrt{s}$. Imposing uniformity on the system when interpreting this superparamagnetic response results in postulating the presence of both AF-fluctuations in the presence of competing ferromagnetic ones \cite{Gegenwart2005} as being intrinsic to the electronic interactions. Our re-analysis \cite{Bretana2023} of the data on the stoichiometric compounds CeRu$_2$Si$_2$, CeCu$_6$, and YbRh$_2$Si$_2$ and their qualitative agreement with the cluster scenario demonstrated that this postulate was unnecessary. It also sidesteps the inherent difficulty -- when trying to understand the experiments in terms of a uniform scenario -- of having to uncover a mechanism through which the FM fluctuations become subleading with respect to the AF fluctuations from the AF-ordered phase.

\subsection{Differences with Griffiths phase}

At first glance it might seem that our cluster scenario is a mere relabeling of the Griffiths phase \cite{CastroNeto2000,Vojta2006} in which fluctuations of the order parameter produce rare regions of order that have a disproportionate influence on the response of the system. However, the two scenarios are fundamentally different and distinct. In the cluster scenario close to the percolation transition, very large isolated clusters form with a large superspin. However, these fluctuations are caused by the ionic motion and their appearence and disappearance occur over timescales much slower than the electronic degrees of freedom, the part of the system that displays the non-Fermi liquid behavior. Second, these rare ordered regions do not play an oversized role in the response of the system: their superspins are so large that they are most likely to be frozen out. The main part of the response to, for instance, an ac-probe originates from the vast collection of smaller isolated clusters \cite{Bretana2023}.  As such, Griffiths phase physics does not appear to be directly relevant to criticality in the cluster scenario, even though it may still play some minor part as discussed in the next section.

\section{Doniach phasediagram updated}

\begin{figure}
\begin{center} 
\includegraphics*[viewport=0 0 560 505,width=85mm,clip]{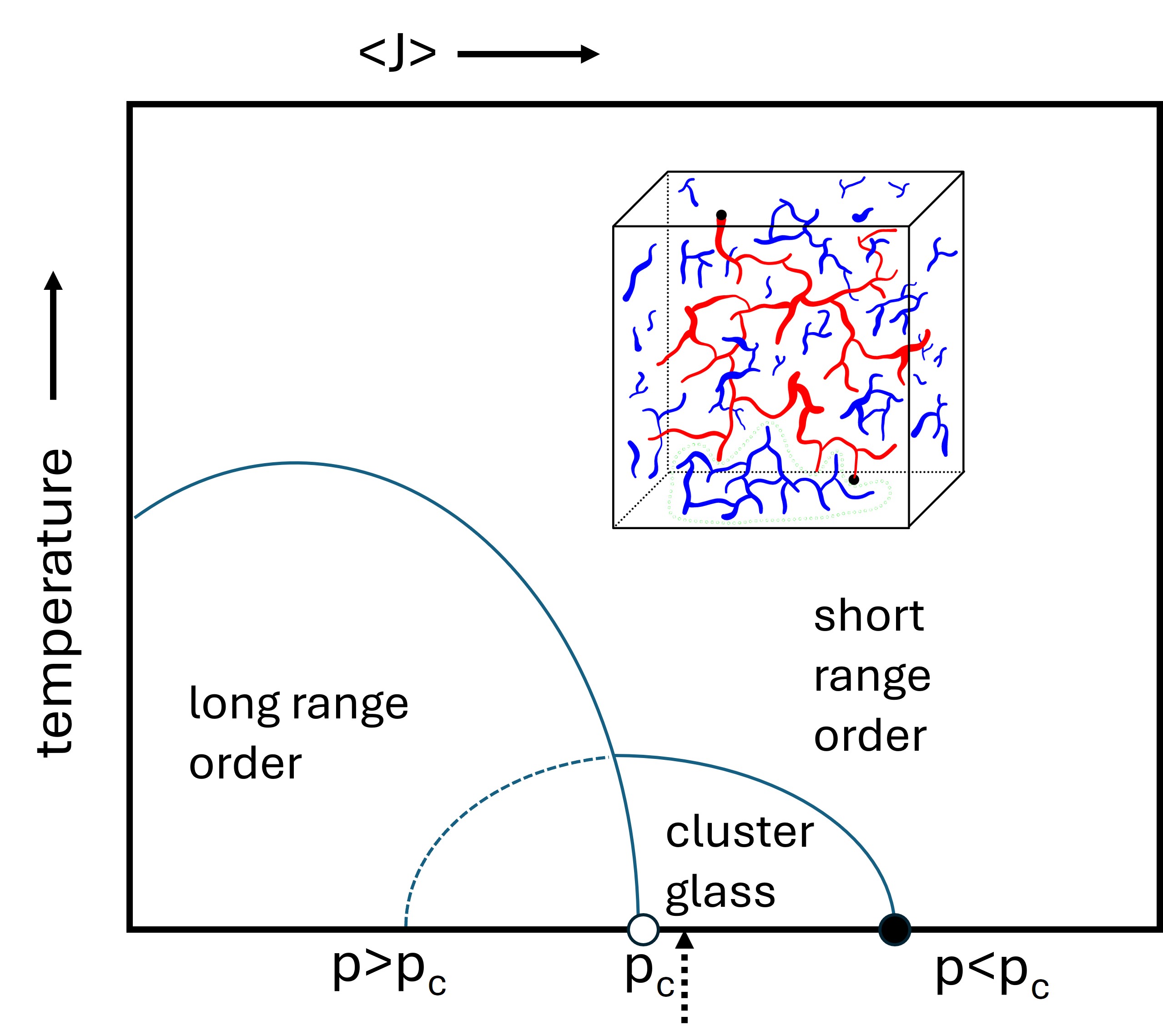}
\end{center}
	\caption{An updated sketch of the Doniach phasediagram. The vertical axis is the temperature of the system, the horizontal axis is a measure of both the average coupling $<J>$ between the conduction electrons and the magnetic moments as well as the average  fraction $p$ of magnetic moments still present at any moment in time. The filled circle on the $T$= 0 axis corresponds to the state of the system where the isolated clusters can no longer line up with each other at any temperature. The shapes in the cube indicate the isolated, ordered clusters (in blue), as well as the lattice spanning cluster (in red). The open circle on the $T$= 0 axis indicates the percolation threshold where the infinite cluster fragments into smaller clusters. For higher occupancies, the infinite cluster survives and its moments will line up below a $p$-dependent transition temperature $T_N$, producing long range order. Above $T_N$ the moments on the infinite cluster remain disordered. For lower occupancies, the infinite cluster breaks up into smaller clusters that will acquire a superspin. Under the dome marked 'cluster glass', the long-range part of the RKKY interaction aligns their superspins (see text). This dome can be destroyed with the application of a magnetic field or through chemical pressure. The dotted arrow indicates where YbRh$_2$Si$_2$ is (likely) located. Note that there exists a range of coupling strength, for $p > p_c$, where we can expect to see two magnetic transitions on cooling, corresponding to ordering on the infinite cluster and to a superspin alignment transition.} 
	\label{Doniach}
\end{figure}

The Doniach phasediagram \cite{Doniach1977} is at the heart of our conceptual understanding of quantum phase transitions. This diagram requires modifications in the cluster scenario, and we have proposed an updated diagram before in the literature \cite{Fayfar2021a}. However, based on our discussion on YbRh$_2$Si$_2$ and experimental results on cobalt substitution for rhodium \cite{Klingner2011}, we believe additional modifications are necessary. We sketch the new phasediagram in Fig. \ref{Doniach}. 

On cooling, magnetic moments will be shielded according to the instantaneous Kondo distribution. As dicussed, this leads to a percolation scenario where isolated clusters with superspins coexist with the lattice spanning, infinite cluster of uncorrelated moments as disordering fluctuations on this clusters do not require a minimum activation energy in view of its large size. The percolation threshold $p_c$ (open circle) marks the boundary where the cluster either survives down to $T$= 0, or breaks up into isolated clusters. If it survives, then at some temperature the moments on this cluster will line up and long range order materializes. If it breaks up, a situation we believe describes YbRh$_2$Si$_2$, then we end up with a system consisting of weakly interacting superspins. The residual interaction between the clusters can be mediated
by the heavy quasiparticles associated with the Kondo-screened moments in the
surrounding lattice. The resulting effective coupling is of RKKY type and
therefore sensitive to perturbations that modify the low-energy spin
susceptibility of the heavy electrons. An applied magnetic field tends to
polarize the heavy quasiparticles and reduces their spin susceptibility,
thereby weakening the effective inter-cluster coupling. As a consequence, the
temperature scale associated with the collective freezing or ordering of the
superspins is expected to decrease with increasing magnetic field. A similar
trend is expected under Ge substitution: replacing Si by Ge
expands the lattice (negative chemical pressure), which in Yb-based Kondo
systems generally reduces the Yb$^{2+}$ valency in favor of Yb$^{3+}$. Within this picture, both an applied magnetic
field and Ge substitution act to suppress the low-temperature ordering scale
associated with the superspin degrees of freedom.

Another interesting aspect is the situation where we have both a collection of superspins and a surviving infinite cluster. In YbRh$_2$Si$_2$ the system can be driven into this direction by applying hydrostatic pressure, or by chemical pressure through Co-substitution on Rh-sites. Upon applying minimal pressure, driving the system closer to $p_c$, we would first expect to see an increase in freezing temperature for the cluster glass. Then, upon increasing pressure we expect to see two magnetic transitions. In experiments with Co-substitution \cite{Klingner2011}, two transitions have indeed been observed. While we have used YbRh$_2$Si$_2$ as our guide, we anticipate that this phasediagram is valid for all quantum critical compounds. Note that in this phase diagram, the quantum critical point that has been studied experimentally in YbRh$_2$Si$_2$ corresponds to the zero ordering temperature of the cluster glass, not to the percolation threshold.

Lastly, it is possible that at some points in time the instantaneous Kondo distribution will be such that an unusually large cluster can materialize, even in the phase where the infinite cluster is still present and the superspins of the  isolated clusters have not lined up with each other. In essence, we would have a Griffiths phase \cite{CastroNeto2000,Vojta2006} at this fleeting moment in time. However, we emphasize the word fleeting: this rare supercluster might wield a very strong influence on its surroundings, but it will only survive on a ps-timescale and it is not expected that such eventualities will be the determining factor in the averaged response under any experimental conditions. 

\section{Conclusions and Outlook}
We have shown that the periodic potential ansatz, which lies at the foundation of modern condensed-matter theory, fails in a fundamental way in quantum critical heavy-fermion systems. This failure does not originate from disorder, chemical substitution, or extrinsic inhomogeneity, but arises inevitably from quantum zero-point motion of the lattice combined with the sensitivity of the Kondo coupling to interatomic separations. The resulting broad distribution of local Kondo temperatures destroys spatial uniformity, even in perfectly stoichiometric crystals. Low-energy electronic states are no longer extended over the full crystal, but become spatially textured, reflecting the underlying distribution of local couplings. This intrinsic nonuniformity provides a natural microscopic origin for the emergence of magnetic clusters and superspin degrees of freedom, leading to a percolative description of quantum critical behavior. Within this framework, the diverse low-temperature phenomenology of heavy-fermion systems — including non-Fermi-liquid thermodynamics, anomalous magnetic response, and multiple low-energy scales — follows in a unified manner. This raises the question of how existing theoretical frameworks may be repaired or extended in a controlled manner since the approach of dismissing the effects of the chaotic electronic environment no longer appears to be a good starting point.

A first and minimal repair strategy is to abandon the requirement of strict translational invariance at low energies while retaining it as a useful organizing principle at higher energies. In practice, this suggests treating the electronic system as composed of spatially fluctuating regions characterized by locally well-defined energy scales, with observables emerging from statistical averages over these regions. Such an approach preserves much of the conceptual machinery of band theory and Fermi-liquid theory while acknowledging that their domain of validity is limited by the absence of self-averaging at low energies. Thus, we need to let go of the current paradigm in correlated electron systems of the ground state being uniform.

A second strategy is to explicitly incorporate percolative degrees of freedom into low-energy theories. The present results indicate that the essential physics near quantum criticality is governed not by uniform order-parameter fluctuations, but by the connectivity and breakup of the lattice-spanning cluster of unshielded moments. Effective field theories that track the evolution of this connectivity, rather than only local order parameters, may therefore provide a more faithful description of quantum critical phenomena in these systems.

Our results suggest that future theoretical efforts should focus on developing controlled approaches to electronic structure in intrinsically nonuniform quantum materials. This includes extensions of density functional theory and dynamical mean-field methods that relax the assumption of strict periodicity and allow for slow spatial variations of local energy scales, as well as new computational frameworks capable of capturing the coexistence of coherent quasiparticles with emergent cluster physics. Perhaps DFT is most-easily adaptable: one can artificially enlarge the unit cell and impose random fluctuations on the electronic interactions within this new cell, with the standard periodic ansatz still in place between the enlarged unit cells. After solving, this process can be repeated and averaged results can be obtained that can be compared to experiments. In addition, by playing around with the enlargement factor, it would be possible to ascertain how intrinsic disorder influences the measurable quantities.

Lastly, we note that the uncertainty associated with zero-point motion will not affect every Kondo system. For instance, Kondo isolators such as Ce$_3$Bi$_4$Pt$_3$ and SmB$_6$ have average Kondo temperatures of $\sim$100 K, well above the lower temperature range where they exhibit insulating behavior \cite{Dzero2016}. Having said that, the distribution of Kondo temperatures can still shed light on phenomena that take place well below the average Kondo temperature. For instance, URu$_2$Si$_2$ undergoes a hidden order transition at 17.5 K \cite{Palstra1985,Mydosh2011}, well below the average Kondo scale. At 17.5 K, a fraction of 0.11 of the moments remains unshielded (Fig. \ref{kondoplot}), and given the two conduction electrons per uranium ion, this implies that the system still has 0.22 Rln2 in entropy to shed. This is exactly the amount that is lost at the hidden order transition \cite{Mydosh2011}. This realization that Kondo shielding was incomplete enabled a reinterpretation of this transition by mapping it onto the dynamics of superfluid helium \cite{Montfrooij2024}, demonstrating that the hidden order transition was associated with an occupancy driven critical dampening of propagating (extended crystal electric field) singlet excitations.

In summary, once the periodic potential approximation fails, the conceptual foundations of single-particle coherence must be reexamined. In a translationally invariant system, electronic eigenstates can be labeled by crystal momentum, and quasiparticles form long-lived excitations that propagate coherently across the entire sample. In strongly correlated electron systems, however, the spatial variations of the local Kondo scale destroy the global coherence required for strict Bloch labeling.


\begin{thebibliography}{10}
\bibitem{explain1}
In our simulation we encoded the moments as having energy $\pm \mu_{\text{eff}} \mu_B B$; for an effective spin half system $\mu_{\text{eff}}$ differs by 2 from the measured $g$-values.
\end{thebibliography}
\end{document}